
\input phyzzx.tex
\PHYSREV


\def\BH{black hole}

\def\beh{\beta_{h}}

\def\gabe{\Gamma_{\beta}}
\def\pano{\par \noindent}
\def\sch{Schwarzschild}
\def\ee{entanglement entropy}

\def\rext{\rho_{ext}}
\def\rint{\rho_{int}}
\def\rh{r_{bh}}
\def\rbo{R_{box}}
\def\text{Tr_{+}}
\def\tint{Tr_{-}}
\def\sent{S_{ent}}

\def\fim{\phi_{-}}
\def\fimp{\phi_{-}^{\prime}}
\def\fip{\phi_{+}}
\def\fipp{\phi_{+}^{\prime}}
\def\omela{\omega_{\lambda}}

\def\erla{R_{\lambda}}
\def\cald{{\cal D}}

\def\rh{r_{bh}}

\date{Jan. 96}
\titlepage
\title{Problems in Black Hole Entropy interpretation\foot{
Talk presented at the ``Fourth Italian-Korean meeting on Relativistic 
Astrophysics'', Rome - Gran Sasso - Pescara, July 9-15, 1995}}
\author{Stefano LIBERATI\foot{ E-mail: Liberati@neumann.sissa.it}}
\address{Scuola Internazionale Superiore di Studi Avanzati, Via Beirut 2-4, 
34013 Trieste, Italy \pano
and I.N.F.N., Sezione di Roma \pano
\pano
\pano
SISSA Ref. 14/96/A}
\endpage

\abstract{In this work some proposals 
for black hole entropy interpretation are exposed and investigated.
In particular I will firstly consider the so called ``\ee " 
interpretation,
\REF\tH{`t Hooft G.\journal Nucl. Phys. &B256  (1985)727.}
in the framework of the brick wall model [\tH ], and the divergence 
problem arising in 
the one loop calculations of various thermodynamical 
quantities, like entropy, internal energy and heat capacity. It is shown that
the assumption of equality of entanglement entropy and Bekenstein-Hawking 
one appears to give 
inconsistent results. 
\REF\BL{Belgiorno F., Liberati S.- {\em Divergences problem in black hole 
brick-wall model} - Preprint gr-qc/9503022.} 
These will be a starting point for a different interpretation of
black hole entropy based on peculiar topological structures of manifolds with 
``intrinsic" 
thermodynamical features. It is possible to show an exact relation between 
\BH\  gravitational 
entropy (tree level contribution in path integral approach) and topology of 
these Euclidean 
space-times. The expression for the Euler characteristic, through the 
Gauss-Bonnet integral, 
and the one for entropy for gravitational instantons are proposed in a form 
which makes the relation between these self evident. Using this relations I 
shall 
propose a generalization of Bekenstein-Hawking entropy in which the former 
and 
Euler characteristic are related in the equation: $S=\chi A/8$. 
\REF\LP{Liberati S., Pollifrone G. - {\em Entropy and
topology for manifolds  with boundaries} - Preprint hep-th/9509093.}
The results, quoted above, are more largely exposed in previous works 
[\BL ,\LP ].
Finally I'll try to expose some conclusions and hypotheses about possible 
further development of this research.}
\endpage

\chapter{Introduction: Black Holes Thermodynamics}

At the beginning of the seventies the research on \BH\ physics achieved a 
series of
theoretical results that brought to an elegant and impressive formulation of 
some General
Relativity laws as thermodynamical ones
(Bardeen, Bekenstein, Carter, Christodoulou, Hawking, 
Ruffini\REF\BCH{Bardeen J.M., Carter B., Hawking S.W.\journal Comm. 
Math. Phys. &31 (1973)161.}
\REF\Chd{Christodoulou D.\journal Phys. Rev. Lett. &25 (1970)1596.}
\REF\CR{Christodoulou D., Ruffini R.\journal Phys. Rev. &D4 (1971)3552.}
[\BCH , \Chd , \CR ]). It was found that for each classical thermodynamical law
it is possible to  formulate a  correspondent one for \BH s. These results are
resumed in the frame below \pano
{\bf 0th Law}:  The surface gravity $\kappa$ is constant on the event
horizon  of a stationary
\BH\
\pano
{\bf 1st Law}:  The mass of a \BH\ is bounded to its area $A$, surface gravity 
$\kappa$ and angular
momentum $J$ by the relation ($ \Omega = $ angular velocity of the \BH ) 
$$
M={{\kappa }\over {2 \pi}} A+\Omega J
$$
\pano
{\bf 2nd Law}:  The area of \BH\ event horizon can never decrease for 
processes satisfying
the weak energy condition\foot{Consider a space-time with curvature tensor 
$R_{ab}$ and matter
described by a stress-energy tensor $T_{ab}$. 
If $\xi^{a}$ is a null or timelike 
vector then from
Einstein equations one has
$$
 R_{ab}\xi^{a}\xi^{b}=8\pi \left [ T_{ab}-
{{1}\over{2}}Tg_{ab} \right ] \xi^{a}\xi^{b}=8\pi \left [ 
T_{ab}\xi^{a}\xi^{b}-
{{1}\over{2}}T \right ]
$$

It is possible to see $T_{ab}\xi^{a}\xi^{b}$ as the energy density measured by 
an observer with
4-velocity $\xi^{a}$ along the geodesic. 
It is commonly accepted that for classical 
matter this energy density has not to 
be negative definite 
that is $T_{ab}\xi^{a}\xi^{b}\geq 0$ for all the $\xi^{a}$ timelike. 
This is called the {\em weak energy condition}. 
It may also be requested that the stress energy tensor of matter cannot be so 
big and negative to
render the 1st member of the preceding relation negative. One can then 
impose the {\em strong 
energy condition}
$$
 T_{ab}\xi^{a}\xi^{b}\geq -{{1}\over{2} }T
 $$
for all the vectors $\xi^{a}$ timelike.}
$$
\delta A \geq 0
$$
\pano
{\bf 3rd Law}:  It is not possible to render the surface gravity of a \BH\ 
null through physical
transformations.
\pano

Although these laws soon appeared as a strong hint towards a 
thermodynamical behaviour
of \BH , the picture became fully consistent only when Hawking found his 
famous results about \BH\ 
radiation by an application of quantum field theory in curved space. 
It is in fact 
impossible to define a 
temperature for classical \BH s, thus it is not even reasonable to talk 
about entropy in this case.
Hawking suggested that, due to the polarization of the vacuum in proximity 
of the \BH\ horizon, 
there is a flux of radiation flowing out towards infinity. He 
demonstrated this by an ingenious 
derivation based on Bogoliubov coefficients
technique \REF\Haww{Hawking S.W.\journal Comm. Math. Phys. &43 (1975)199.} 
[\Haww ].
By the first law and the Hawking temperature one finds the Bekenstein-Hawking
entropy of \BH\ to be one quarter of its area [\Haww]. 

Thermodynamical aspects of black holes
appear more evident in Euclidean  
path-integral approach\REF\GH{Gibbons G.W.,
Hawking S.W.\journal Phys. Rev. &D15 (1977) 2752.} [\GH ]. 
Considering the generating functional of 
the Euclidean theory with the action equal to the
Einstein-Hilbert one plus the matter contribution,  
in a semiclassical approach the tree level contribution is due only to 
gravitational part. Instantons are non singular solutions of the classical 
equations in 4-dimensional Euclidean space. For a wide class of \BH  
space-times, metrics that extremize the Euclidean action are gravitational 
instantons if one removes the conical singularity at the horizon.
This forces to fix a period for the imaginary time. It is well known that 
Euclidean quantum field theory with periodic imaginary time 
is equivalent to a finite temperature quantum field theory in 
pseudoeuclidean $(-,+,+,+)$ space-time, the
temperature being the inverse of imaginary time period. Thermodynamics 
appears in this way as a
request of consistence of quantum field theory on \BH\ space-times, or 
better, on space-times with Killing horizon.

\section{Interpretation problems for black hole entropy.}

As it often happens in the history of science, the discovery of such a 
complex structure opened 
a whole set of new questions to which complete answers still lack. 
Some questions are in order:\pano
1	Which dynamical degrees of freedom could be associated with BH 
entropy?
\pano
2	Is there any information loss in \BH\ dynamics ? 
\pano
3 How does General Relativity know about \BH\ thermodynamics ?

I will now expose more extensively these points which appear as the main 
problematics opened by the research in
quantum aspects of \BH . 
\pano
1 The problem of a dynamical origin for \BH\ entropy is the effort to achieve a 
statistical
mechanics explanation of it. This means to give an interpretation of horizon 
thermodynamics in a
``familiar'' way, in the sense that it could be seen as related to dynamical 
degrees of freedom
associated with \BH\ nature.
\pano
2 The information loss is related to the fact that \BH\ 
evaporation by mean of emission of a thermal particle spectrum, appears to
produce a destruction of quantum information by converting pure state in 
mixed ones. That is hardly
acceptable by the great part of physicists because it would imply a 
non-unitary evolution
of quantum states in presence of strong gravitational fields.
\pano
3 The last point is, I believe, the most important one.  
It in fact appears as the main
question one has to answer in order to understand the real nature of horizon 
thermodynamics.
As we have just seen, the four laws of \BH\ thermodynamics have been 
formulated some years before
Hawking's discovery of quantum radiation. Most of these appear in fact as 
General Relativity theorems of \BH\
dynamics and so are already ``encoded" at a classical level. But in
spite of this, the  consistence of this frame
is achieved only at the quantum stage by the introduction of Hawking 
radiation. 
How can geometry know about quantum matter behaviour is still an 
unsolved question. We shall see
how the attempt to explain tree level (classical) gravitational contribution 
as due to matter one
(one loop) appears to fail. 

In the last twenty years many authors tried to give answers to these
questions. What is common to most part of them is the conviction that all
these problems are related and that a clear comprehension of \BH\ entropy 
origin
would be a great achievement in order to solve them. So we shall start
considering the first point of \BH\ entropy, $S=A/4$, statistical
interpretation. Here I shall only quote some major proposals:
\pano
1	Bekenstein\REF\Be{Bekenstein J.D.\journal Phys. Rev. &D9 
(1974)3292.}[\Be ] -
The \BH\ entropy can be seen as a $S=\ln W$ where W is the number of possible
microscopical  configurations of 
astrophysical body that generate the same BH (relation to the ``no hair" 
theorem).
\pano
2	York\REF\Yo{York J.W.\journal Phys. Rev. &D28 (1983)2929.} [\Yo ] - 
The dynamical degrees of freedom at the origin of \BH\ entropy are 
identified with BH  
``quasi normal modes". 
\pano
3 Wald\REF\Wa{Wald R.M.\journal Phys. Rev. &D48 (1993)3427.} [\Wa ] - Black 
hole
entropy is identified with the Noether charge, associated to a diffeomorphism
invariant theory, in presence of bifurcate Killing horizons\foot{A Killing
horizon is null hypersurface whose null generators are orbits of a Killing
vector field. In General Relativity it has been demonstrated that the event
horizon of a stationary \BH\ is always a Killing horizon. If the generators of
the horizon are geodetically complete to the past (and the surface gravity of the
\BH\ is different from zero) then it contains a 2-dimensional (in four
dimensional space-times) space-like cross section $B$ on which the Killing 
vector is null. 
$B$ is called ``surface of bifurcation", is a fixed point for
the Killing flow and on it the Killing vector vanishes. $B$ lies at the
intersection of the two hypersurfaces (past and future) forming the complete
horizon. A bifurcate Killing horizon is an horizon with a  ``surface of
bifurcation" $B$.}.\pano 
4 Srednicki\REF\Sr{Srednicki M.\journal Phys. Rev. Lett. &71 (1993)666.} 
[\Sr ]- 
Bombelli, Koul, Lee, Sorkin\REF\BKLS{Bombelli L., Koul R.K., Lee J. and 
Sorkin
R.D.\journal Phys. Rev. &D34 (1986)373.} [\BKLS ]
- Frolov, Novikov\REF\FN{Frolov V.P., Novikov I.\journal Phys. Rev. &D48 
(1993)4545.} [ \FN ] - 
\BH\ entropy is generated by dynamical degree of freedom, excited at a 
certain time,  associated to the matter in BH interior near the horizon 
through non-causal 
correlation (EPR) with external matter. The ignorance of an observer outside 
the \BH\ about the
modes inside the horizon is associated with a so called entanglement 
entropy which is identified
with Bekenstein-Hawking one.
\pano
5	Hawking\REF\HH{Hawking S.W., Horowitz G.T., Ross S.F. - {\em Entropy, 
area and black hole pairs} - Preprint gr-qc/9409013.} [\HH ]- 
Black hole entropy has a
topological origin.
The topological structure of space-time determines the presence of 
gravitational entropy 
\pano
The first two lines of research appear at the moment not successful due to 
their appeal to a count
over the entire life of \BH . This point has been stigmatized by a recent
Gedanken experiment by
Frolov and Novikov\REF\FNW{Frolov V., Novikov I.\journal Phys. Rev. &D48
(1993)1607.} [\FNW ].  In this work a wormhole is used as a device to 
explore the backside of the  horizon, the behaviour
found seems to show a deep relation between some sort of dynamical degrees 
of freedom living behind
the horizon and \BH\ entropy. The third proposal (that of Noether charge) 
shows a deep link between rescaling properties of the (gravitational plus 
matter) action and the presence of thermodynamical behaviour in presence 
of bifurcate Killing horizons. 
Unfortunately, although very impressive, it lacks in giving
a properly statistical interpretation of \BH\ entropy. In this sense it is more
a way of recovering Bekenstein-Hawking results that casts a new light on the
nature of the problem than an interpretative frame. 
So in the rest of this work we are going to study the last two proposals in
order to understand what the right answer to our questions might be.

\chapter{Entanglement Entropy}

Let us consider a global Hilbert space ${\cal H}$ composed of two 
uncorrelated ones ${\cal H}_{1}$, 
${\cal H}_{2}$.
$$
{\cal H}={\cal H}_{1} \otimes {\cal H}_{2}
$$                  
A general state on {\cal H} can be described as a 
linear superposition of states on the two Hilbert
spaces
$$
| \psi \rangle = \sum_{a,b} \psi(a,b) |a \rangle | b\rangle
\eqn\gs
$$
One can define a global density matrix
$$
\rho(a,a',b,b')=\psi(a,b) \psi^{*} (a',b')
$$
The reduced density matrix for the subsystem $a$ is given by
$$
\rho(a,a')=\sum_{b,b'}\psi(a,b) \psi^{*} (a',b')
\eqn\rd
$$
Note that even if the general state \gs\ defined on the global Hilbert space 
is a pure one, the 
form of the
reduced density matrix \rd\ shows that the corresponding state defined only 
in one subspace is a 
mixed one.
This corresponds to an information loss that can be properly described by 
defining an entropy 
associated to
mixed states which is null for pure ones, that is the so called von Neumann 
entropy
$$
S(a)=-Tr(\rho_{a} \ln \rho_{a})
$$                        
Let us consider now the same problem in \BH\ case.

Let us consider a stationary \BH\ and define $\hat{\rho}^{iniz}$ the density 
matrix describing, in
Heisenberg representation, the initial state of quantum matter 
propagating on its background. 
For an external observer the system consist of two parts: \BH\ and radiation 
outside of it. By defining a spacelike hypersurface we can consider 
quantum modes of radiation at a given time so that they
can be separated in external and internal to the \BH . 
The state for external radiation is obtainable 
from $\hat{\rho}^{iniz}$ by tracing on all the state of matter 
inside the event horizon
and so inaccessible to the external observer
$$
\hat{\rho}^{rad}=Tr^{inv}\hat{\rho}^{iniz}
$$
For a \BH\ alone this density matrix would describe its Hawking radiation at 
infinity.
We can also define the density matrix for the \BH\ state
$$
\rho^{BH}=Tr^{vis}\hat{\rho}^{iniz}
$$
where one now performs the trace on external degrees of freedom.
From this matrix is it possible to find the related von Neumann entropy
$$
S^{BH}=-Tr^{inv}(\hat{\rho}^{BH} \ln \hat{\rho}^{BH})
\eqn\Enta
$$
This is exactly what is called ``entanglement entropy".
It is important to note that this definition is invariant in the sense that 
independent changes of
definitions of vacuum for ``external" and ``internal" states do not change the 
value of $S^{BH}$.
The calculation of entanglement entropy can be resumed in few steps
\pano
- Introduction of null-like or space-like hypersurface (achronal ``slice") on 
which perform matter quantization.\pano
- 2nd quantization of matter field on BH background.\pano
- Definition of the global ground state.\pano
- Trace of the global density matrix on internal or external states.\pano
- Computation of \ee\ entropy as von Neumann entropy of reduced density 
matrix.\pano


Calculations on \ee\ were performed by various authors in a wide class of
situations, in flat spaces as in curved ones 
[\Sr ], [\BKLS ], [\FN ].
In spite of this we find two common points that appear
as proper characteristics of \ee .\pano
1 Entanglement entropy is proportional to the ``division" plane
(for BH is the event horizon).\pano
2 Entanglement entropy is always divergent on division plane due
to the presence of modes of arbitrary high modes near the horizon.
The divergence form is general and independent on the kind of field.

While the first point is a strong hint towards the identification of \ee\ 
with Bekenstein-Hawking one, the second is a deep problem casting some 
shadows on the real
understanding of the meaning of what the computation of \ee\ is effectively 
probing about
\BH\ physics.

The last years have seen different approaches for the divergence problem
resolution. Roughly they can be summarized in two kind of proposals that of
regularization (t'Hooft; Frolov, Novikov; Barvinsky, Frolov, Zelnikov) and
that of renormalization (Susskind, Uglum; Fursaev, Solodukhin).

The main idea of regularization [\FN ]
\REF\BFZ{Barvinsky A.O., Frolov V.P., Zelnikov A.I.\journal 
Phys. Rev. &D51(1995)1741.} approach is to apply a physical cut-off to 
entropy by
justifying it through the quantum fluctuations (Zitterbewegung) of the event 
horizon.
This cut-off has been estimated by Frolov and Novikov [\FN ] and it is of 
the order of the
Planck length. Remarkably the introduction of such a cut-off gives a value of
\ee\ of the same order of the Bekenstein-Hawking one.

The second approach \REF\susug{Susskind L. and Uglum J.\journal 
Phys. Rev. &D50 
(1994)2700.} is instead based on elimination of divergences through
a renormalization of gravitational 
coupling constant [\susug] and of constants related to second 
order curvature terms\REF\FS{Fursaev D.V., Solodukhin S.N. - 
{\em On one-loop renormalization of black hole entropy} - 
Preprint hep-th/9412020.}[\FS ]. This way appears very interesting for its 
relation to
elementary particle physics but is penalized by its necessity of a 
renormalizable theory of
quantum gravity in order to give exact results.\pano
 
We shall now consider the regularization approach in order to probe its 
consistency.
The first idea of regularization of \ee\ 
was implicitly proposed by `t Hooft in 
1985 [\tH ] as applied to his ``brick wall model". 
In a certain sense cut-off dependent models [\FN,\BFZ] are up 
to date versions of the former. One of the problems `t Hooft proposed in his 
seminal work was the divergence of not only entropy but also of quantum 
matter contribute to internal energy of the \BH , which has to be regularized 
by using the same cut-off one has to introduce for entropy. 
He found that, fixing 
the cut-off in order to obtain $\sent=S_{Bek-Haw}=A/4$, one obtains $U={3 
\over {8}}M$. So matter contribution to internal energy appeared to be a very 
consistent fraction of the black hole mass $M$. As `t Hooft underlined, this is 
a signal for a strong  back-reaction effect, not a good aim for a model 
based on semiclassical
(negligible back-reaction) approximation.\pano 
We shall see that the same problem
is present in Barvinsky, Frolov, Zelnikov (BFZ) model [\BFZ] and that a 
surprising behaviour of heat-capacity is also found. 

\chapter{Entanglement Entropy and BFZ Model}

In BFZ work entropy is computed from global vacuum density 
matrix by tracing over the degrees of freedom of matter outside the \BH. 
In so doing one obtains a mixed state density matrix for matter 
inside the \BH . \pano 
BFZ define the global wave function of the \BH\  as
$$
\Psi=\exp(\Gamma/2) \langle \fim | \exp(-\beta \hat{H}/2) 
| \fip \rangle
\eqn\fdon
$$ 
and
$$
\hat{\rho}=| \Psi \rangle \langle \Psi |
\eqn\den
$$
as the related density matrix. Here $|\fip \rangle$ ($|\fim 
\rangle$) are the 
external (internal) states of matter (a massless scalar field for 
simplicity) 
on the \BH\  fixed background.\pano
Tracing over $|\fip \rangle$ gives the internal density matrix
$$
\eqalign{
\rint(\fimp,\fim)&=\langle \fimp|\hat{\rho}|\fim \rangle 
\cr 
&=\int \cald \fip \Psi^{*}(\fimp,\fip)\Psi(\fim,\fip)\cr
&=\exp(\gabe)\langle \fimp|\exp(-\beta\hat{H})| \fim 
\rangle. \cr} 
\eqn\din
$$
Entanglement entropy associated to this reduced density matrix is
$$
S_{ent}=-Tr_{int}(\rho_{int} \ln \rho_{int})
$$
Here $\gabe$ is a normalization factor fixed in order to obtain $ 
tr \rho=1 $,
but it also corresponds to the 1 loop effective action
$$
\eqalign{
\gabe &=-\ln \left [ \int \cald \fim \langle \fim | \exp(-
\beta \hat{H}) |\fim
\rangle \right ] \cr
&=-{1\over{2}} \ln \det \left [{{\delta(x-y)}\over{2(\cosh 
\beta
\omega -1)}} \right ]\cr}
\eqn\gam
$$
where $\hat{\omega}$ is the operator associated with the 
frequency of field
modes.\pano
\REF\Bee{Bekenstein J.D. - {\em Do we understand black hole entropy ?} - 
Preprint gr-qc/9409015} 
BFZ calculate the \ee\ (in WKB approximation) as the trace over the internal 
modes of $-\rint \ln \rint $, so their calculation is relative to 
the internal degrees of freedom. Instead, in the common 
definition of \ee, one usually refers to the trace over the 
external degrees of freedom of $- \rext \ln \rext $. In the 
following, I shall assume\foot{For a demonstration of this assumption 
see [\Bee ].} that, given the symmetry existing in BFZ study between 
internal 
and external variables, the two definitions of \ee\ coincide.
Moreover it is possible to demonstrate [\BL ] that all other 
thermodynamical quantities, as internal energy and heat capacity, share 
the same property. 

From the definitions one has $$
\eqalign{
\rext &=\tint | \Psi \rangle \langle \Psi |=\int \cald \fim
\Psi^{*}(\fipp,\fim) \Psi(\fim,\fip)=\cr    
&=\int \cald \fim \langle \fipp| \exp
\left ( - {{\beta \hat{H}}\over{2}} \right ) | \fim \rangle 
\langle \fim | \exp
\left ( -{{\beta \hat{H}}\over{2}} \right ) | \fip \rangle \exp 
(\Gamma_{ext})
\cr
\rint &=\text | \Psi \rangle \langle \Psi |=\int \cald \fip
\Psi^{*}(\fimp,\fip) \Psi(\fip,\fim)=\cr    
&=\int \cald \fip \langle \fimp| \exp
\left ( - {{\beta \hat{H}}\over{2}} \right ) | \fip \rangle 
\langle \fip | \exp
\left (-{{\beta \hat{H}}\over{2}} \right ) | \fim \rangle 
\exp(\Gamma_{int})\cr}
\eqn\den
$$

For $\gabe$, from \gam , one obtains (using where we used the property 
$\ln \det A= Tr \ln A$)\pano
$$
\gabe=\int dx 
\left [ \ln \left ( 2 \sinh {{\beta \hat{\omega}}\over{2}} \right 
) \delta(x-y)
\right]_{\bf{y}=\bf{x}} 
\eqn\ugbfz
$$
Following BFZ, one can calculate the expression below by expanding all the 
functions $\phi(x)$ in terms of eigenfunctions $R_{\lambda}(x)$ of the 
operator $\hat{\omega}$
$$
\eqalign{
\phi(x)&=\sum_{\lambda} \phi_{\lambda} 
R_{\lambda}(x)\cr
\hat{\omega}^{2} \erla(x)&=\omela^{2}\erla(x)\cr
\delta(x-y)&=\sum_{\lambda} g^{00} g^{1/2} \erla(x) 
\erla(y) \cr}
\eqn\eigen
$$
where $\sum_{\lambda}$ denotes the sum over all quantum 
numbers, $g^{00}$ is the timelike component of the metric 
tensor and 
$g=\det g_{\mu
\nu}= g^{00}\det g^{ab}$ (a,b,$\ldots = 1,2,3$).\pano
Hence 
$$
\gabe=\int_{2M}^{\rbo} dr {r^{2}\over{(r-
2M)}}\int_{0}^{\infty} 
\sum_{l=0}^{\infty} d\omega (2l+1) R^{2}_{\lambda 
\omega}(r) \gamma (\beta 
\omega) 
\eqn\gammabeta
$$
where
$$
\gamma(\beta \omela)={\beta \over{2}}\omega_{\lambda}+
\ln(1- e^{-\beta \omela})
\eqn\muga
$$
and where $R_{\lambda \omega}(r)$ are the radial 
eigenfunctions. 
We are interested in the behaviour of $\gabe$ near the 
horizon; 
using BFZ result
$$
\sum_{l=0}^{\infty} (2l+1) R^{2}_{\lambda \omega}(r) \sim 
{4\over{\pi}} \omega^{2} {M\over{r-2M}}
\eqn\near
$$
one gets 
$$
\gabe \sim {{4M}\over{\pi}} \int_{2M}^{r_{box}} dr
{{r^{3}}\over{(r-2M)^{2}}}\int_{0}^{+\infty} d\omega 
\omega^{2} 
\gamma (\beta \omega)
\eqn\ganear
$$
where $r_{box}$ is the radius of the box in which we have to 
put the \BH\ to
regularize infrared divergences.\pano
To compute the second integral one has to subtract the 
zero--point term from \muga .
Finally one finds the following leading term near the horizon
$$
\gabe=\beta F(\beta)\sim -{{32 \pi^{3} M^{4}}\over{45}} 
{1\over{\beta^{3}}} {1\over{h}}
\eqn\gaho
$$
where the cut--off is defined as $h\equiv Inf(r-2 M)$.\pano
From the free energy \gaho , it is possible to find the other 
thermodynamical
quantities (not only entropy but also internal energy $U$ and heat 
capacity $c$) by using the well 
known relations between free energy and these ones in canonical ensemble.
So one obtains 
$$\eqalign{
S&\sim {{128 \pi^{3} M^{4}}\over{45}} {1\over{\beta^{3}}} 
{1\over{h}}\cr
U&\sim {{32 \pi^{3} M^{4}}\over{15}} {1\over{\beta^{4}}} 
{1\over{h}}\cr
c&\sim {{128 \pi^{3} M^{4}}\over{15}} 
{1\over{\beta^{3}}}{1\over{h}}\cr} 
\eqn\suc
$$
Rewriting the above formulas in terms of a proper distance 
cut--off
$$
\epsilon \sim 2  \sqrt{\rh h} \Leftrightarrow h\sim 
{\epsilon^{2} 
\over{4 \rh}}
\eqn\proper
$$
we find for $F,\ U,\ S\ {\rm and}\ c$ at the Hawking 
temperature $\beta_{H}={{1}\over{8 \pi M}}$
$$\eqalign{
F(\beh)&\simeq -{M \over{720 \pi}} {1\over{\epsilon^{2}}}\cr
S(\beh)&\simeq {2M^2 \over{45}} {1\over{\epsilon^{2}}}\cr
U(\beh)&\simeq  {M \over{240 \pi}} 
{1\over{\epsilon^{2}}}\cr
c(\beh)& \simeq {{4 M^{2}}\over{30 \epsilon^{2}}}\cr}
\eqn\final
$$
The entropy in \final\ is exactly the same than in BFZ. 

\section{Interpretative Problems}

At this point a brief clarification seems necessary about the 
interpretation one would give to the quantities just found.
The above divergences for the entropy and the other 
thermodynamical quantities requires a renormalization 
scheme or a brick-wall cut-off. 
The standard position consists in identifying the black hole 
entropy with the leading divergent regularized term
$$
S_{bh}\equiv S_{radiation,\ leading}. 
\eqn\bwa
$$ 
But what about the regularized terms for the other 
thermodynamical quantities? 
The cut-off present in \final\ is the same for all the 
thermodynamical quantities so we have to fix the same value 
of $\epsilon$ for all of them. We shall see that the 
values so obtained for $F$, $U$, $c$ are very different from the classical 
\BH\ ones: this means that a straightforward identification between e.g. 
$E_{bh}$ and $E_{radiation,\ leading}$ doesn't seem possible. 
The same kind of problem, even worse, exists for the specific 
heat as we'll see further on.\pano
The identification of Bekenstein-Hawking entropy with the 
entanglement one generates a problem of 
interpretation of classical (tree level) entropy due to gravity in 
the path-integral approach. 
The first aim of entanglement approach is to explain 
all \BH\ entropy as 
dynamical matter entropy. 
The matter leading term is not a new one-loop contribution 
to be added to the tree level one. So it appears as a necessary 
complement of this
program to give a clear explanation for ignoring the presence of the 
tree level contribution of gravity. As a matter of fact, in 
literature this problem appears 
to be often ignored or gone around. We can quote in this sense 
only a work 
by Jacobson \REF\Jac{Jacobson T.- {\em Black hole entropy and 
induced 
gravity} - Preprint gr-qc/9404039.} [\Jac ]. Here I'd like to put in 
evidence that this crucial point is the same as the third question we 
encountered in section 1. The explanation one can give for ignoring 
tree level in \ee\ approach require an answer to the problem of relation 
between classical and quantum aspects in \BH\ thermodynamics.

\REF\fur{Fursaev D.V.\journal Mod. Phys. Lett. &A10 (1995)649.}
\REF\solo{Solodukhin S.N.\journal Phys. Rev. &D51 (1995)609.}
\REF\solod{Solodukhin S.N.\journal Phys. Rev. &D51 
(1995)618.} 
\REF\fursolo{Fursaev D.V. and Solodukhin S.N.- {\em On 
One-Loop Renormalization 
of Black Hole Entropy} - Preprint hep-th/9412020.}
\REF\froa{Frolov V.P.- Why the Entropy of a Black Hole is 
A/4? - Preprint Alberta-Thy-22-94, gr-qc/9406037.}\par 
Following the most part of papers on the same problem 
[\BFZ,\susug,\tH,\fur,\solo,\solod,\fursolo,\froa] 
we can try to check if the identification of Bekenstein-Hawking 
entropy with the 
entanglement one gives self-consistent results at the level of the 
other thermodynamical quantities. My discussion is here limited 
to 
the brick-wall regularization of the divergences 
[\BFZ,\tH,\froa].

\section{Free Energy, Internal Energy and Heat Capacity}

The cut-off fixing necessary to obtain the required value

$S_{ent}=S_{Bek-Haw}=A/4$ is
$$
\epsilon^{2}={{1}\over{90 \pi}}
\eqn\cuto
$$
this brings to the following values for free energy and internal 
energy
$$\eqalign{
F&=-{{1}\over{8}}M\cr
U&={{3}\over{8}}M.\cr}
\eqn\udef
$$
The results in \udef\ are the same obtained by t'Hooft in his 
pioneering 
paper [\tH] and exactly the same are found [\BL ] if one calculates $U$ 
and $F$ with 
heat kernel expansion truncated to the first De Witt 
coefficient in the optical metric
\REF\DK{Dowker J.S., Kennedy G.\journal Journ. Phys. &A11 (78)895.}
[\DK ].
So, for the internal 
energy, the identification of the brick-wall 
value with the tree level one ($M$ the mass of the \BH ) seems impossible; 
on the other hand, it does not seem 
possible to understand the radiation term as a 
perturbative contribution to the black hole tree level one. 
It happens in fact that the quantum contribution is of the same order of 
the classical one.
Moreover there would exist a further problem with the special 
form of the internal energy radiation contribution, the horizon 
contribution being not of the form expected for a massless gas.

I shall now check the behaviour of \BH\ heat capacity in 
brick wall model.\pano
For heat capacity, imposing again the cut-off value \cuto , one 
obtains from
\final\ 
\foot{Also in this case, it is possible to obtain the same value 
from `t Hooft results [\tH] with 
a simple computation.}
$$
c=+12\pi M^{2}
\eqn\cfin
$$
It is important to note that this is a positive value and bigger in 
module than the classical well-known result
$$
c_{class}=-8 \pi M^{2}
\eqn\ccla
$$
So, if we accept the brick-wall model plus entanglement entropy 
frame as
dynamical explanation of \BH\ entropy, we find, in the most 
naive 
interpretation of \cfin, that \BH s are 
stabilized by one loop contribution of matter.

\chapter{Possible explanations and proposal}

The results we have just found sound like a ``warning bell" for proposal of 
entanglement
entropy. The conclusions one can draw from them might be rather radical. 
In fact if we 
accept the results \cfin\ for \BH\ heat capacity they seem to imply a 
``stabilization" 
of the black hole by quantum matter\foot{It is well known that a negative 
heat capacity 
(like the \BH\  classical one) is characteristic of thermodynamical instable 
system}.
Another possibility is that back reaction is never negligible in study of 
quantum effect on
\BH\ backgrounds.\pano
The hypothesis that it is wrong to impose \ee\ equal 
to the 
Bekenstein-Hawking one appears more realistic. 
These considerations bring to consider more deeply the fifth proposal for 
\BH\ entropy explanation we encountered in section 1. In this 
interpretative frame Bekenstein-Hawking entropy is seen as related 
to topology structure of \BH\ manifolds.

Recent works by Hawking, Horowitz and Ross 
\REF\HHi{Hawking S.W., 
Horowitz G.T., Ross S.F.- 
{\em Entropy, Area and black hole pairs} - Preprint gr-qc/9409013.} 
\REF\HHii{Hawking S.W., Horowitz G.T.- {\em The 
gravitational 
hamiltonian, 
action, entropy and surface terms} - Preprint gr-qc/9501014.}
[\HHi ,\HHii ]
have demonstrated that the usual Bekenstein-Hawking law for 
\BH\ entropy 
fails in 
the case of extreme \BH s. For these kinds of object we have a 
null entropy in 
spite of a non null area of the event horizon. These authors 
observed that this
change in extreme case in respect of non-extreme one is mainly 
due to the 
different nature of event horizon in the former.
One in fact finds that, in these cases, the presence of the event 
horizon is not 
associated with a non-trivial topology of space-time.
Euler characteristic is in fact zero (not two) for this kind of 
\BH s. 
This radical difference in extreme \BH\ physics seems a strong 
hints 
towards 
a point of view that particular case of \BH\ solutions 
(for example 
extreme Reissner-N\"ordstrom \BH\ is ``just" the case 
$Q^{2}=M^{2}$ of the 
general solution) is to be considered a rather different object from
the non-extreme one. 
Nevertheless it is possible to understand extreme case as a 
particular case 
of \BH\ without requiring a limitation of \BH\ 
thermodynamics laws.
The guiding idea (originally proposed by Gibbons and Hawking 
\REF\GHI{Gibbons G.W., Hawking S.W.\journal Commun. Math. Phys. &66 (1979) 
291.} [\GHI ])
is that thermodynamical features of space-times like the \sch\ 
one are 
explainable as an effect due to their non-trivial topological 
structure and above all to the nature of their boundaries.
In particular Euler characteristic and entropy 
have the same 
dependence on the boundaries of the manifold and 
we will relate 
them in a general formula. This relation (although 
demonstrated only for a 
certain class of metrics) would be valid for every compact 
manifold on which 
Gauss-Bonnet theorem can be extended. 

\chapter{Euler characteristic and manifold structure}

The Gauss-Bonnet theorem proves that it is possible to obtain 
the Euler 
characteristic of a 4-dimensional compact Riemannian manifold 
$M$ without 
boundaries by the volume integral of the 4-dimensional metric 
curvature
$$
S_{GB}={{1}\over{32 \pi^{2}}} \int_{M} \epsilon_{abcd} 
R^{ab}\wedge R^{cd}
$$
with $R$ bound to the spin-connections $\omega$ of the 
manifold by the
relations
$$
R^{a}_{b}=d\omega^{a}_{b}+\omega^{a}_{c}\wedge \omega^{c}_{b}
$$
Chern 
\REF\CHI{Chern S.\journal Annals of Math. &45 (1944)747.}
\REF\CHII{Chern S.\journal Annals of Math. &46 (1945)674.} [\CHI ,\CHII ]
showed that the differential n-form $\Omega$ of Gauss-Bonnet integral
$$
\Omega={{(-1)^{p}}\over{2^{2p}\pi^{p}p!}} \epsilon_{a_{1} 
\ldots a_{p}} R^{a_{1} a_{2}} \wedge \ldots \wedge 
R^{a_{2p+1} a_{2p}}
$$
defined on $M^{n}$ can be defined on a manifold $M^{2n-1}$ 
which is the 
image
of $M^{n}$ through the flux of its unitary vector field. 
Then he was able to express $\Omega$ as the exterior 
derivative of a
differential $n-1$-form in $M^{2n-1}$
$$
\Omega=d\Pi
$$
He also demonstrated that the original $\Omega$ integral on 
$M^{n}$ can be
performed on a submanifold $V^{n}$ of $M^{2n-1}$ whose boundaries 
are the set of singular points of the unitary vector field previously 
cited.
By Stoke's theorem we then obtain
$$
S^{vol}_{GB}=\int_{M^{n}}\Omega=\int_{V^{n}}\Omega=
\int_{\partial V^{n}}\Pi
$$
For manifold with boundaries this formula can be 
generalized 
\REF\EGH{Eguchi T., Gilkey P.B., Hanson A.J.\journal Phys. 
Rep.
&66, 6 (1980).} [\EGH ]
$$
S_{GB}=S^{vol}_{GB}+S^{bou}_{GB}=\int_{M^{n}}\Omega-
\int_{\partial M^{n}}\Pi=\int_{\partial V^{n}}\Pi-
\int_{\partial M^{n}}\Pi
\eqn\Sgb
$$
This expression implies that the Euler characteristic of a 
manifold $M^{n}$ 
with boundaries becomes null in the case that its contours 
would be the same 
as
that of the submanifold $V^{n}$ of $M^{2n-1}$.\pano
We shall now use \Sgb\ for \BH\ manifold. We always work 
in Euclidean 
manifolds after imaginary time compactification necessary in 
order to remove 
conical singularities on the horizons.

For non-extreme \BH s the boundaries of the manifold $V$
are set by the extreme values of the range of radius coordinate 
that are 
$r=r_{h}$ and $r=r_{0}=\infty$.
The physical manifold $M$ instead has just one boundary at 
infinity because,
after removal of conical singularity, the \BH\ horizon 
$r=r_{h}$ is not a 
border of space-time. 
So
$$
\Sgb=\int_{r_{0}}\Pi -\int_{r_{h}} \Pi -\int_{r_{0}} \Pi=-\int_{r_{h}} \Pi
$$
It is possible to use this formula to calculate Euler number and 
for example
in Schwarzschild case one correctly
obtains $\chi_{euler}=2$ which is the expected value for 
$S^{2}\times R^{2}$
topology.\pano

For extreme \BH s the boundaries of the manifold $V$ 
are the same as the one for the ordinary case $r=r_{h}$ and
$r=r_{0}=\infty$. 
On the other hand the physical manifold $M$ has now two 
boundaries at 
infinity
represented by the usual spatial infinity $r=r_{0}=\infty$ and by 
the horizon 
$r=r_{h}$.
In fact, in this case the time-affine Killing vector 
has a set of fixed
points only at infinity (it becomes null only asymptotically at 
infinity, in this
sense one says that the \BH\ horizon for extreme \BH\ is at 
infinity). 
So
$$
\Sgb=\int_{r_{0}}\Pi -\int_{r_{h}} \Pi -\int_{r_{0}} \Pi 
+\int_{r_{h}} \Pi=0
$$
This shows that for extreme \BH\ the Euler characteristic is 
always null.

\chapter{Entropy for manifolds with boundaries}

We will follow the definition of \BH\ entropy adopted by 
Kallosh, Ortin, 
Peet \REF\KOP{Kallosh R., Ortin T., Peet A.\journal Phys. 
Rev. &D47 (1993)5400.} [\KOP ].\pano

Let us consider a thermodynamical system with conserved 
charges $C_{i}$ and
relative potentials $\mu_{i}$ so that we work in grand 
ensemble.
$$
\eqalign {
Z &={\rm{Tr}}\; e^{-(\beta H -\mu_{i} C_{i})} \cr
Z &=e^{-W}\cr
W &=E-TS-\mu_{i} C_{i} \cr }
$$
we obtain
$$
S=\beta(E-\mu_{i}C_{i})+\ln Z
$$
Gibbons-Hawking demonstrated that at the tree level
$$
\eqalign{
Z& \sim e^{-I_{E}}\cr
I_{E}&={{1}\over{16 \pi}} \int_{M}(-
R+L_{matter})+{{1}\over{8\pi}}
\int_{\partial M} \left [ K \right ] \cr }
$$
Here $I_{E}$ is the ``on-shell'' Euclidean action.\pano
In calculating $Z$, and hence $I_{E}$, it is important to correctly 
evaluate the
boundaries of our manifold M.\pano 
For no-extreme \BH\ we have just one boundary at infinity 
$r_{0}
\rightarrow \infty $ (after the removal of conical singularity, the 
metric is regular
on the horizon $r=r_{h}$).\pano
For extreme \BH\ we have a drastic change in boundaries 
structure.
Metrics do not present conical singularity so we cannot fix 
imaginary time 
value. The horizon is at an infinite distance from the external 
observer and so 
it is like an ``internal" boundary of our space-time (we can say 
that the
coordinate of this internal boundary is $r_{b}$).

In order to determine $S$ we also have to compute 
$\beta(E-\mu_{i}C_{i})$.\pano
From Gibbons-Hawking [\GH] we know that for two fixed 
hypersurfaces at 
$\tau=cost$ 
($\tau=$imaginary time), 
$\tau_{1}$ e $\tau_{2}$, one has
$$
\langle \tau_{1} | \tau_{2} \rangle =e^{-(\tau_{2}-\tau_{1})(E-
\mu_{i} C_{i})}
\approx e^{-I_{E}}
$$
In this case it is necessary to understand that the time-affine 
Killing vector
$\partial/\partial \tau$ has two sets of fixed points, one at 
infinity and the other on the horizon. 
So an hypersurface at $\tau=cost$ has two 
boundaries in
corresponding to these sets, independently of the position of 
horizon (which
can be at infinity for extreme \BH\ ).

So one obtains\foot{Note that, for metrics under our 
consideration, 
$ V_{bulk}=M_{bulk} $  so the bulk part of the entropy always 
cancels also for
metrics which are not Ricci-flat (as de Sitter case). All the 
entropy depends
on boundary values of extrinsic curvature.}
$$
\eqalign{
S&=\beta(E-\mu_{i} C_{i}) +\ln{Z}=\cr
&={I_{E}}^{\infty}_{r_{h}}-{I_{E}}^{\infty}_{r_{boun}}\cr
&={{1}\over{8 \pi}} \left (\int_{\partial V} 
[K] - \int_{\partial M} [K ] \right)=\cr
&={{1}\over{8 \pi}} \left ( \int_{r_{0}} [K]- \int_{r_{h}} [K]-\int_{r_{0}} 
[K]+\int_{r_{b}} [K]
\right )
\cr}
\eqn\entr
$$ 
The deep similarity between \entr\ and \Sgb is self-evident.

In the case of extreme \BH , we don't have an internal boundary for $M$ 
and so we don't have  
$r_{boun}$ in \entr . Hence
$$
S={{1}\over{8 \pi}} \left [ \int_{\infty} [K]- \int_{r_{h}}[K] 
- \int_{\infty} [K] \right]=-{{1}\over{8 \pi}} \int_{r_{h}}[K]=A/4
$$

On the contrary, in the case of extreme \BH\  the horizon is at infinity and 
$M$ has two boundaries in 
$r=\infty$
and $r_{b}=r_{h}$
$$          
S={{1}\over{8 \pi}} \left [ \int_{\infty} [K]- \int_{r_{h}}[K]  
- \int_{\infty} [K] +\int_{r_{h}}[K]  \right] =0
$$

Some comments on derivation \entr\ are in order.
It was in fact derived in a grand ensemble but for 
extreme
\BH\ there is no conical singularity so there is no $\beta$ 
fixing and consequently no intrinsic thermodynamics of the manifold. 
We conjecture 
that the correct
procedure we have to follow is exactly the inverse. The last 
line of \entr\ is
the general expression of entropy for manifold with 
boundaries. The lack of
intrinsic thermodynamics is deducible from \entr\ by 
consideration of 
boundary
structure. It is not possible to fix $\beta$ because boundary 
changes in 
extreme
case, not the contrary. Thus \entr\ is generalizable to a large
class of Riemannian manifolds with boundaries and the 
similarity in boundary
dependence with Gauss-Bonnet integral is a strong hint towards 
the evidence of a  
link between
entropy and topology for gravitational instantons.
 
\chapter{General case: spherically symmetric metrics}

In order to find a general relation linking Euler characteristic of the 
manifold 
to the gravitational entropy, we consider, for simplicity, metrics of the 
form $$
ds^{2}=-e^{2U(r)}dt^{2}+e^{2U(r)}dr^{2}+R^{2}(r)d^{2}\Omega
\eqn\metr
$$ 
On having
$$
\eqalign{
A &=4 \pi R^2 (r_h)\cr
\beta &=4\pi((e^{2U})'_{r=r_h})^{-1}\cr
S &= \left.{{\beta R}\over{2}} [(U'R+2R')e^{U}-
{{2R}\over{r}}]e^{U}\right|_{r=r_h}\cr
\chi &= \left.{{\beta}\over{2 \pi}}(2U' e^{2U})(1-
e^{2U}R'^{2})\right|_{r=r_h}\cr}
$$
one finds
$$
\eqalign{
S &=2 \pi \chi (2U' e^{2U})^{-1}_{r=r_h} (1-e^{2U}R'^{2})^{-1}_{r=r_h} 
{{R}\over{2}} 
\left. [(U'R+2R')e^{U}-{{2R}\over{r}}]e^{U}\right|_{r=r_h}=\cr
&= \pi \chi [(e^{2U})'-R'^{2} e^{2U}(e^{2U})']^{-1}_{r=r_h} [ {{\beta 
R}\over{2}}
( e^{2U})'+2R'e^{2U}-{{\beta 2R}\over{r_h}}e^{U}]_{r=r_h}\cr}
$$
being $\left. e^{2U}\right|_{r=r_h}=0$. 

Finally one has
$$
\eqalign{
S &= \pi \chi R(r_h) \left \{ \left [ (e^{2U})' \right]^{-1} \right
\}_{r=r_h}=\cr &= {{\pi \chi R^{2}(r_h)}\over{2}}={{\chi (4 \pi
R^{2}(r_h))}\over{8}}= {{\chi A}\over{8}} \cr}
\eqn\SC
$$
Such a relation points out the deep link between the gravitational 
entropy and the topological structure of the manifold.

Note that this formula gives the correct result for the extreme \BH\ cases 
(for which usual Bekenstein-Hawking formula fails) and that it also has a 
general 
validity for all the metrics of the form \metr\ \foot{The extension of \SC\
to metrics of more general form is at the moment under investigation 
\REF\LPP{Liberati S., Pollifrone G. -{\em Work in progress}.}[\LPP ].}.

\chapter{Conclusions and Perspectives}

From this analysis, we deduce that the interpretation of \BH\  
entropy as \ee\ brings to problematic
results for internal energy and heat capacity.
These are not identifiable with their background counterparts 
so these have to be seen (differently from the entropy 
situation) as quantum correction to the corresponding tree 
level 
gravitational terms.\pano
The internal energy, as `t Hooft remarked [\tH], 
is of the same order of magnitude of \BH\ mass: at this point 
one must question the applicability of the assumption of the 
negligible back-reaction.
Even if we pass over this problem, we still find that the one 
loop contribute of matter to \BH\ heat capacity is positive and 
so it would stabilize the \BH\ . But we believe it is an 
inconsistent result because 
quantum correction is, for the heat capacity, bigger than its 
background 
counterpart; it could be more plausible 
if we would have quadratic terms in curvature tensor in the gravitational
action, but this is not our 
case.
Our results can be interpreted either as a proof of the inconsistency 
of the
identification of \ee\ and Bekenstein-Hawking one, or as a structural 
``bug" embedded in the brick- wall problem approach to \BH\ 
thermodynamics 
if one ignores 
the back-reaction of matter field on the gravitational back-ground.

On the other side relation \SC\ appears to hold in a wide class of manifolds.
It seems that this formulation of \BH\ entropy sheds new light on the 
behaviour
of the extreme \BH s by interpreting the gravitational entropy 
as a topological effect (in this sense it confirms Hawking's position). \pano
Unfortunately it appears rather difficult to find a dynamical 
explanation of this (topological) entropy. 
We conjecture that this  
relation of the gravitational entropy with boundary structure of the 
space-time is in a certain sense a hint towards an interpretation based on 
dynamical degrees of 
freedom associated to the vacuum states in non-trivial topologies.\pano 
In particular one may propose a deep relation between deformation due to 
topological changes in
zero-point modes of quantum fields and thermal effect on black holes 
space-times
\REF\BLC{Belgiorno F., 
Liberati S. - {\em Work in progress}.}[\BLC ].  \par
Moreover this approach 
seems to require, as a consistency  condition, a point of view
towards gravitational action which gives a possible answer to the problem we 
exposed previously about
explanation of presence of \BH\ laws already in General Relativity. 
\REF\Sak{Sakharov A.D.\journal Soviet Physics - Doklady 
&12, (1968)1040.}
In fact it would force, in some sense, an interpretation of gravitational 
action as an effective one someway induced by a more fundamental quantum 
matter level\foot{Perhaps in a way similar to Sakharov ``induced 
gravity" program [\Sak ]}. In this sense it would be  
meaningless to speak about ``pure gravitational" action and paradoxes 
encountered in identification of
tree level contribution with 1 loop one might be solved.
\REF\Jc{Jacobson T.\journal Phys. Rev. Lett. &75 (1995)1260.} 
Moreover this appears consistent with the recent Jacobson hypothesis [\Jc ] 
about the 
interpretation of General Relativity as the thermodynamical limit of a 
more fundamental theory. 

As a matter of fact 
investigation about
thermal nature of horizon characterized space-times appears as a crucial step 
toward a deeper
comprehension of the essence of gravitation.

\ACK{I am grateful to F.Belgiorno and G.Pollifrone for useful 
remarks and discussions}.

\endpage\refout
\end